\documentclass[11pt]{article}

\usepackage[utf8]{inputenc}
\usepackage[T1]{fontenc}
\usepackage[margin=1in]{geometry}
\usepackage{amsmath,amssymb,amsthm,mathtools}
\usepackage{bm}
\usepackage{hyperref}
\usepackage{enumitem}
\usepackage{microtype}
\usepackage{graphicx}
\graphicspath{{figures/}}
\usepackage{booktabs}
\usepackage{placeins}
\usepackage{xcolor}

\hypersetup{
  colorlinks=true,
  linkcolor=blue,
  citecolor=blue,
  urlcolor=blue
}

\numberwithin{equation}{section}

\newcommand{\lr}{\lambda_r}
\newcommand{\dd}{\Delta}
\newcommand{\eps}{\varepsilon}

\newcommand{\DS}{\mathrm{DS}}

\title{%
A Cusp Criterion for a Local Phase-Space Bottleneck\\
in a Minimal Adiabatic Marcus Hamiltonian}

\author{
    Stephen Wiggins \\[1ex]
    \small Hetao Institute of Mathematics and Interdisciplinary Sciences, \\
    \small Shenzhen, Guangdong Province, China \\[1ex]
    \small School of Mathematics, University of Bristol, \\
    \small Bristol, BS8 1UG, United Kingdom
}
\date{\small Author accepted manuscript\\[0.3em]
Published in \emph{Chemical Physics} \textbf{613} (2027), 113517.\\
\url{https://doi.org/10.1016/j.chemphys.2026.113517}\\[0.3em]
\footnotesize This manuscript version is made available under the\\
\href{https://creativecommons.org/licenses/by-nc-nd/4.0/}{CC-BY-NC-ND 4.0 license}.}

\begin{document}

\maketitle

\begin{abstract}
Marcus--Hush theory describes electron transfer through reorganization energies, driving forces, and electronic coupling, but it does not by itself determine when an adiabatic model possesses a local Hamiltonian bottleneck. We study a minimal asymmetric two-degree-of-freedom lower-sheet Hamiltonian with an exactly separable harmonic transverse mode. For this conservative model, we establish a necessary-and-sufficient root count for the stationary points and recover the known cusp boundary for disappearance of the lower-sheet double well. We then interpret that boundary in Hamiltonian phase space: inside the cusp, the lower sheet has an index-one saddle and the associated local transition-state structures---an unstable periodic-orbit NHIM, its stable and unstable manifolds, and an attached dividing surface---exist; outside the cusp, this local saddle-mediated structure is absent. The result applies to this conservative adiabatic potential-energy Hamiltonian. It does not describe solvent free-energy dynamics, bath-induced mode coupling, or nonadiabatic transfer, and it does not predict Marcus rates for realistic condensed-phase electron-transfer systems.
\end{abstract}

\section{Introduction}

Marcus--Hush theory describes electron transfer through diabatic free-energy surfaces, reorganization energies, driving forces, and electronic couplings \cite{Marcus1956,Hush1961,NewtonSutin1984,MarcusSutin1985}. In this description, the diabatic surfaces represent uncoupled electronic states, and the reorganization energy measures the energy cost of nuclear rearrangement. This framework underlies descriptions of thermal and optical electron transfer, intervalence charge transfer, and localization in mixed-valence compounds \cite{RobinDay1968,BrunschwigCreutzSutin2002,HeckmannLambert2012,PartheyKaupp2014}.

A useful qualitative motivation is intramolecular electron transfer in class II mixed-valence systems, for which two localized diabatic valence states, a nuclear reorganization coordinate, and an electronic coupling provide a standard minimal description \cite{RobinDay1968,PiephoKrauszSchatz1978,BrunschwigCreutzSutin2002,HeckmannLambert2012,PartheyKaupp2014}. The present equal-curvature model is not fitted to a particular compound and is not intended to provide a quantitative Robin--Day classification. It is used only to ask the following geometric question within a minimal two-state setting: when does the lower adiabatic potential support an index-one saddle and the associated local Hamiltonian phase-space bottleneck?

In multidimensional Hamiltonian systems, transition states are phase space objects. Near an index-one saddle they are organized by normally hyperbolic invariant manifolds (NHIMs), together with their stable and unstable manifolds and attached dividing surfaces possessing a local no-recrossing property \cite{Wigner1938,UzerEtAl2002,WaalkensSchubertWiggins2008,EzraWaalkensWiggins2009}. In two degrees of freedom, the NHIM is an unstable periodic orbit; in higher dimensions, it becomes a higher-dimensional invariant sphere. The local no-recrossing property of the attached dividing surface provides the basis for \emph{local} transition state theory near the bottleneck.

For electron transfer, however, several mathematically distinct settings coexist and should be kept separate. In one line of development, the high-dimensional solvent problem is projected onto a single scalar variable, often the energy gap, leading to reduced diabatic free-energy curves or a one-dimensional potential of mean force \cite{NewtonSutin1984,MarcusSutin1985}. In another, the environment is treated as a thermal bath, leading to stochastic, non-Hamiltonian dissipative descriptions such as the Zusman and Sumi--Marcus theories \cite{Zusman1980,SumiMarcus1986}. In still another, the nonadiabatic weak-coupling regime is treated by mixed quantum-classical, mapping, or surface-hopping approaches, including the quantum-classical Liouville equation and MASH \cite{Kapral2006,LawrenceMannouchRichardson2024}. The present paper does not attempt to merge these viewpoints. It isolates only the conservative, adiabatic, lower-sheet Hamiltonian regime.

Starting from a minimal asymmetric two-state model with one nuclear reorganization coordinate and one exactly separable transverse oscillator, we ask when the lower adiabatic potential possesses an index-one saddle. The closed-form boundary at which the lower-sheet double well disappears has previously appeared in the electron-transfer literature \cite{JungCao2002}. The direct root and stability classification given here identifies the same boundary as the necessary-and-sufficient condition, within this conservative model, for a saddle--centre equilibrium and hence for the associated \emph{local} NHIM/UPO and dividing-surface construction. The transverse mode is a separable spectator oscillator, so the model has no bath-induced mode coupling, intermode energy redistribution, or nonintegrable transport.

This analysis is related to work on Hamiltonian saddle-node bifurcation and reaction dynamics \cite{GarciaGarridoNaikWiggins2020}. The present problem is more specific: the bifurcation condition is derived directly from the lower adiabatic Marcus sheet itself, and the resulting cusp gives a Marcus-type criterion for the birth or loss of the local bottleneck. The upper adiabatic sheet, dissipative solvent dynamics, and nonadiabatic transitions remain relevant in other regimes. Here we determine when a minimal adiabatic Marcus Hamiltonian supports a lower-sheet phase space transition state.

\section{Diabatic Hamiltonian model and lower adiabatic surface}

We use a minimal two-state vibronic Hamiltonian of the displaced-harmonic type familiar from Marcus--Hush and mixed-valence models \cite{PiephoKrauszSchatz1978,Nitzan2006,MayKuhn2011}. The electronic degree of freedom is represented by the two-dimensional donor/acceptor diabatic state space; the variables $y$ and $z$ below are nuclear coordinates, not electronic position coordinates. The coordinate dependence of the diagonal diabatic energies couples the electronic state to nuclear reorganization, while $\dd$ couples the two electronic diabatic states. The model is intentionally finite-dimensional and does not explicitly represent all molecular or solvent coordinates.

In keeping with the adiabatic focus of this paper, the role of the full
two-state model is only to generate the lower adiabatic potential on which
the subsequent classical dynamics takes place. Consider the diabatic
Hamiltonian acting on the two-dimensional electronic diabatic-state space,
\begin{equation}
H(y,z,p_y,p_z)
=
\frac{1}{2}(p_y^2+p_z^2)\,I_2
+
\begin{pmatrix}
U_D(y,z) & \dd\\[1mm]
\dd & U_A(y,z)
\end{pmatrix},
\label{eq:diabatic-H}
\end{equation}
where \(I_2\) denotes the \(2\times2\) identity matrix. The donor and
acceptor diabatic potentials are
\begin{equation}
U_D(y,z)
=
\frac{1}{2}\Omega^2(y+y_0)^2
+
\frac{1}{2}\omega^2 z^2,
\qquad
U_A(y,z)
=
\frac{1}{2}\Omega^2(y-y_0)^2
+
\eps
+
\frac{1}{2}\omega^2 z^2 .
\label{eq:diabatic-potentials}
\end{equation}
Here $y$ is a \emph{nuclear reorganization coordinate} associated with electron transfer, $z$ is a nuclear transverse spectator mode, $\dd>0$ is the electronic coupling between the donor and acceptor diabatic states, and $\eps$ is their diabatic energy asymmetry. The transferred electron is represented by the two-state electronic basis in Eq.~\eqref{eq:diabatic-H}; it is not represented by an additional classical coordinate.

The associated one-coordinate harmonic Marcus reorganization
energy is
\begin{equation}
\lr = 2\Omega^2 y_0^2 .
\label{eq:reorg-energy}
\end{equation}
Indeed, the two diabatic minima are separated by \(2y_0\), so this is the
standard displaced-parabola value
\(\lambda=(1/2)\Omega^2(2y_0)^2\) for equal-curvature Marcus surfaces
\cite{MarcusSutin1985,Nitzan2006,MayKuhn2011}.

The adiabatic surfaces are obtained by diagonalizing the \(2\times2\)
diabatic potential matrix:
\begin{equation}
V_\pm(y,z)
=
\frac{U_D(y,z)+U_A(y,z)}{2}
\pm
\sqrt{
\left(
\frac{U_D(y,z)-U_A(y,z)}{2}
\right)^2
+
\dd^2
}.
\label{eq:adiabatic-surfaces}
\end{equation}
For each fixed nuclear configuration $(y,z)$, $V_+(y,z)$ and $V_-(y,z)$ are the two eigenvalues of the coupled electronic potential matrix in Eq.~\eqref{eq:diabatic-H}. They are therefore the upper and lower adiabatic potential-energy surfaces of the two-state model, not additional hypothetical states introduced independently of it. Propagating classical nuclei only on $V_-$ is an adiabatic approximation: transitions to $V_+$ are neglected and would require a nonadiabatic treatment.

A direct calculation gives
\begin{equation}
\frac{U_D+U_A}{2}
=
\frac{1}{2}\Omega^2(y^2+y_0^2)
+
\frac{\eps}{2}
+
\frac{1}{2}\omega^2 z^2,
\qquad
\frac{U_D-U_A}{2}
=
\Omega^2 y_0 y
-
\frac{\eps}{2}.
\label{eq:diabatic-average-difference}
\end{equation}
Consequently, the lower adiabatic surface is
\begin{equation}
V_-(y,z)
=
\frac{1}{2}\Omega^2(y^2+y_0^2)
+
\frac{\eps}{2}
+
\frac{1}{2}\omega^2 z^2
-
\sqrt{
\left(
\Omega^2 y_0 y-\frac{\eps}{2}
\right)^2
+
\dd^2
}.
\label{eq:lower-sheet}
\end{equation}

The adiabatic Hamiltonian studied in this paper is therefore the scalar
lower-sheet Hamiltonian
\begin{equation}
H_-(y,z,p_y,p_z)
=
\frac{1}{2}(p_y^2+p_z^2)
+
V_-(y,z).
\label{eq:lower-sheet-Hamiltonian}
\end{equation}
All subsequent analysis concerns the local dynamics of the conservative lower-sheet Hamiltonian near a critical point of $V_-$. The functions $U_D$, $U_A$, and $V_-$ are finite-dimensional potential-energy functions in this model; they are not solvent free-energy surfaces or potentials of mean force. Accordingly, the cusp condition derived below does not determine Marcus rate constants, the validity of thermal Marcus kinetics, or condensed-phase transition-state theory. It determines only whether this constructed lower adiabatic potential possesses an index-one saddle and the associated local Hamiltonian bottleneck geometry. The restriction to $V_-$ assumes adiabatic nuclear motion and excludes transitions to the upper sheet.

\section{Equilibria, stability, and the asymmetric cusp}

Because the $z$-dependence of $V_-$ remains purely harmonic, the critical-point problem reduces to a one-dimensional nonlinear equation in the electron-transfer coordinate $y$. Throughout this paper, we use \emph{critical point} for a point of the potential at which its first derivatives vanish, and reserve \emph{equilibrium point} for a fixed point of Hamilton's equations. In the symmetric case ($\eps=0$), this analysis leads to the standard threshold condition $\dd<\lr/2$, a well-known result in classical electron-transfer theory \cite{Nitzan2006,MayKuhn2011}. In the asymmetric case, however, the saddle is shifted away from the origin. While the intuitive loss of the barrier under strong asymmetry or coupling is chemically familiar, its boundary is less commonly highlighted in the context of phase space transport. Here we show that the existence of the saddle is governed by a cusp-type bifurcation in parameter space, and we give a complete account of the corresponding equilibria of the Hamiltonian flow and their linear stability.

\subsection{Critical points of the lower adiabatic surface}

To find the critical points of the lower adiabatic surface $V_-$, we differentiate Eq.~\eqref{eq:lower-sheet} with respect to $z$ and $y$. For the transverse mode,
\begin{equation}
\frac{\partial V_-}{\partial z}=\omega^2 z,
\label{eq:dzV}
\end{equation}
so every critical point must lie at $z=0$. For the electron-transfer coordinate, the derivative is
\begin{equation}
\frac{\partial V_-}{\partial y}
=
\Omega^2 y
-
\frac{\Omega^2 y_0\left(\Omega^2 y_0 y-\eps/2\right)}
{\sqrt{\left(\Omega^2 y_0 y-\eps/2\right)^2+\dd^2}}.
\label{eq:dyV}
\end{equation}
Setting this equal to zero yields the location of the critical points $(y,0)$:
\begin{equation}
y
=
y_0\,
\frac{\Omega^2 y_0 y-\eps/2}
{\sqrt{\left(\Omega^2 y_0 y-\eps/2\right)^2+\dd^2}}.
\label{eq:critical-equation-dimensional}
\end{equation}

It is convenient to introduce the dimensionless variables
\begin{equation}
x=\frac{y}{y_0},
\qquad
\beta=\frac{\eps}{\lr},
\qquad
\delta=\frac{2\dd}{\lr}.
\label{eq:dimensionless-variables}
\end{equation}
Substituting these into Eq.~\eqref{eq:critical-equation-dimensional}, the critical-point equation becomes
\begin{equation}
x=\frac{x-\beta}{\sqrt{(x-\beta)^2+\delta^2}},
\label{eq:critical-equation-dimensionless}
\end{equation}
To avoid any reliance on roots introduced by squaring, the complete admissible-root count is given in Appendix~\ref{app:root-count}. The direct argument shows that Eq.~\eqref{eq:critical-equation-dimensionless} has exactly three roots if and only if
\begin{equation}
|\beta|<\bigl(1-\delta^{2/3}\bigr)^{3/2},
\qquad 0<\delta<1,
\label{eq:direct-three-root-condition}
\end{equation}
with the roots ordered minimum--saddle--minimum. On the regular boundary with $0<\delta<1$, there is one simple root and one double root. At the cusp tip $(\beta,\delta)=(0,1)$ there is one triple root at $x=0$; elsewhere outside the three-root region and its regular boundary there is exactly one simple root. For numerical root finding only, squaring Eq.~\eqref{eq:critical-equation-dimensionless} gives
\begin{equation}
(1-x^2)(x-\beta)^2=\delta^2x^2,
\label{eq:quartic-factor-form}
\end{equation}
and hence
\begin{equation}
x^4-2\beta x^3+(\beta^2+\delta^2-1)x^2+2\beta x-\beta^2=0,
\label{eq:quartic-expanded}
\end{equation}
with every numerical candidate checked in the original unsquared equation; the tolerances used are stated in Appendix~\ref{app:numerics}.

\subsection{Lifting critical points to equilibria of the Hamiltonian flow}

The lower-sheet Hamiltonian $H_{-}$ given in \eqref{eq:lower-sheet-Hamiltonian} is of standard kinetic-plus-potential (``natural'') form. For such a Hamiltonian, an equilibrium point of the associated Hamiltonian vector field must satisfy $\partial H_{-}/\partial p = 0$ and $\partial H_{-}/\partial q = 0$. Hence $p_y=p_z=0$ and $(y,z)$ must be a critical point of $V_{-}$. Consequently, every critical point $(y_*,0)$ of the potential naturally lifts to an equilibrium point $(y_*,0,0,0)$ of Hamilton's equations.

\subsection{Linear stability of the equilibria}

The stability of an equilibrium is determined by the linearisation of Hamilton's equations. Around $(y_*,0,0,0)$ one obtains
\[
\frac{d}{dt}
\begin{pmatrix}
\delta y \\ \delta z \\ \delta p_y \\ \delta p_z
\end{pmatrix}
=
\begin{pmatrix}
0 & 0 & 1 & 0 \\
0 & 0 & 0 & 1 \\
-\partial_{yy}V_- & 0 & 0 & 0 \\
0 & -\omega^2 & 0 & 0
\end{pmatrix}
\begin{pmatrix}
\delta y \\ \delta z \\ \delta p_y \\ \delta p_z
\end{pmatrix},
\]
where the derivatives of $V_-$ are evaluated at $(y_*,0)$. The eigenvalues $\mu$ of this block matrix satisfy $\mu^2 = -\lambda$, with $\lambda$ an eigenvalue of the Hessian $D^2V_-(y_*,0)$. Because the Hessian is diagonal,
\[
D^2V_-(y_*,0)=
\begin{pmatrix}
\partial_{yy}V_- & 0 \\[2pt]
0 & \omega^2
\end{pmatrix},
\]
its eigenvalues are $\partial_{yy}V_-$ and $\omega^2$.

Since $\omega^2>0$, the transverse direction always contributes a pair of purely imaginary eigenvalues $\pm i\omega$, corresponding to a centre (elliptic) motion. The stability therefore hinges entirely on the sign of $\partial_{yy}V_-$:
\begin{itemize}
    \item If $\partial_{yy}V_->0$, the equilibrium is \textbf{centre--centre} (elliptic in both degrees of freedom).
    \item If $\partial_{yy}V_-<0$, the equilibrium is \textbf{saddle--centre} (hyperbolic in the reaction coordinate $y$, elliptic in $z$).
\end{itemize}
An index-2 saddle (saddle--saddle equilibrium) cannot occur because the $z$-curvature is strictly positive. On the regular bifurcation boundary, a saddle--centre coalesces with a centre--centre in a Hamiltonian saddle-node bifurcation \cite{GarciaGarridoNaikWiggins2020}. The cusp tip is a higher-order degeneracy, treated separately below. A general discussion of index-$k$ saddles in Hamiltonian systems can be found in the recent monograph by Wiggins, García-Garrido, and Katsanikas~\cite{wiggins2025phase}.

To decide which case actually occurs, we compute the $y$-curvature. From \eqref{eq:dyV},
\begin{equation}
\frac{\partial^2 V_-}{\partial y^2}(y_*,0)
=
\Omega^2
-
\frac{\Omega^4 y_0^2\dd^2}
{\left[\left(\Omega^2 y_0 y_*-\eps/2\right)^2+\dd^2\right]^{3/2}}.
\label{eq:yy-curvature-dimensional}
\end{equation}
In the dimensionless variables, this becomes
\begin{equation}
\frac{1}{\Omega^2}\frac{\partial^2 V_-}{\partial y^2}(y_*,0)
=
1-\frac{\delta^2}{\bigl((x_*-\beta)^2+\delta^2\bigr)^{3/2}}.
\label{eq:yy-curvature-dimensionless}
\end{equation}
For $x_*\ne\beta$, the right-hand side also equals $(x_*^3-\beta)/(x_*-\beta)$. The exceptional stationary point $x_*=\beta=0$ must be evaluated using the nonsingular expression in Eq.~\eqref{eq:yy-curvature-dimensionless}, which gives $1-1/\delta$. Thus a critical point corresponds to a saddle--centre equilibrium exactly when
\begin{equation}
1-\frac{\delta^2}{\bigl((x_*-\beta)^2+\delta^2\bigr)^{3/2}}<0.
\label{eq:saddle-sign-condition}
\end{equation}

\subsection{The asymmetric cusp and its regular saddle-node branches}

On a regular branch of the bifurcation boundary, a saddle--centre equilibrium disappears when it coalesces with a centre--centre equilibrium. This happens when the first and second derivatives of $V_-$ with respect to $y$ vanish simultaneously, i.e., when $\partial_{yy}V_-=0$ at a critical point. At a nonzero coalescence point, combining the critical-point equation with the zero-curvature condition in Eq.~\eqref{eq:yy-curvature-dimensionless} gives $x^3=\beta$. Substituting this into \eqref{eq:quartic-factor-form} yields
\[
(1-x^2)(x-x^3)^2=\delta^2 x^2.
\]
For $x\neq0$, dividing by $x^2$ gives $(1-x^2)^3=\delta^2$, whence $x^2=1-\delta^{2/3}$ and $\beta=x^3$. The bifurcation locus in the $(\beta,\delta)$-plane is therefore
\[
|\beta|=(1-\delta^{2/3})^{3/2},\qquad 0<\delta<1.
\]
Along these regular branches there is a \textbf{Hamiltonian saddle-node bifurcation} of equilibria: a saddle--centre and a centre--centre merge and annihilate. The excluded case $x=0$ requires $\beta=0$ in the critical-point equation and $\delta=1$ in the zero-curvature equation. It is the cusp tip, where all three stationary roots meet and the equilibrium has zero reactive curvature; it is not a regular saddle-node point. Away from the bifurcation boundary, inside the cusp there are three equilibria (two centre--centre, one saddle--centre), whereas outside it only a single centre--centre remains.

In the original dimensional variables, the lower adiabatic surface $V_-$ possesses an index-one saddle (and the Hamiltonian $H_-$ a saddle--centre equilibrium) if and only if
\begin{equation}
\left|\frac{\eps}{\lr}\right|
<
\left(1-\left(\frac{2\dd}{\lr}\right)^{2/3}\right)^{3/2},
\qquad
\frac{2\dd}{\lr}<1.
\label{eq:cusp-criterion-dimensional}
\end{equation}
Equivalently, in the dimensionless $(\beta,\delta)$-plane,
\begin{equation}
|\beta|<(1-\delta^{2/3})^{3/2},
\qquad
0<\delta<1.
\label{eq:cusp-criterion-dimensionless}
\end{equation}
The symmetric limit of Eq.~\eqref{eq:cusp-criterion-dimensional} is $\eps=0$ and $\dd<\lr/2$, the familiar lower-sheet double-well threshold for the equal-curvature adiabatic Marcus--Hush model \cite{Nitzan2006,MayKuhn2011}. In the equivalent form
\begin{equation}
|\eps|^{2/3}+(2\dd)^{2/3}<\lr^{2/3},
\label{eq:cusp-astroid-dimensional}
\end{equation}
the boundary for disappearance of the lower-sheet double well was previously obtained in electron-transfer theory by Jung and Cao \cite{JungCao2002}. We use the word ``cusp'' in the standard bifurcation-theoretic sense of the parameter-space boundary on whose regular branches two critical points coalesce and at whose tip all three meet \cite{Arnold1989}. The contribution made here is the direct admissible-root count and Hamiltonian stability interpretation: inside this boundary the conservative lower-sheet model has one saddle--centre equilibrium and two centre--centre equilibria, whereas outside it has only one centre--centre equilibrium.

\section{Chemical interpretation of the cusp criterion}

The cusp boundary has an established chemical interpretation as the loss of a double-well lower adiabatic potential in an equal-curvature two-state model \cite{JungCao2002}. Here we interpret this boundary in the local Hamiltonian phase space of the conservative model. The interpretation concerns potential-energy geometry; it does not determine solvent free-energy barriers, thermal rates, or compound-specific mixed-valence classifications.

\textbf{Phase space bottlenecks and NHIMs.}  In conventional transition-state theory, especially in its configuration-space presentation, the transition state is often associated with a saddle point on a potential energy surface. However, in multidimensional systems with non-zero kinetic energy, identifying the true dynamical bottleneck requires moving from configuration space to phase space (coordinates and momenta) \cite{Wigner1938}. In phase space, the transition state is not a single point but a codimension-one surface in the energy shell, and the relevant local question is whether trajectories passing through the bottleneck cross that surface exactly once. 

Modern reaction dynamics has established that this multi-dimensional dividing surface is anchored by a Normally Hyperbolic Invariant Manifold (NHIM) \cite{UzerEtAl2002,WaalkensSchubertWiggins2008}.  In a two-degree-of-freedom system, the NHIM is an unstable periodic orbit located above the potential saddle. Its stable and unstable manifolds are invariant tube-like surfaces in the energy shell that organize trajectories approaching and leaving the bottleneck \cite{GarciaGarridoNaikWiggins2020}. Thus, once an index-one saddle is present, the standard NHIM construction supplies the local phase space object to which a locally no-recrossing dividing surface can be attached.

\textbf{Established boundary and phase-space interpretation.} The loss of the lower-sheet double well under increasing electronic coupling or asymmetry is established in adiabatic electron-transfer theory, and the closed-form asymmetric boundary has been reported previously \cite{JungCao2002,Nitzan2006,MayKuhn2011}. For the present conservative separable Hamiltonian, the direct root count and curvature classification show that this same boundary is necessary and sufficient for the existence of an index-one saddle and therefore for the associated \emph{local} NHIM/UPO and dividing-surface geometry. The following consequences refer only to this model:

\begin{enumerate}[label=(\roman*)]
\item \textbf{Effect of electronic coupling in the model.} At fixed asymmetry, increasing $\dd$ increases $\delta=2\dd/\lr$. Once the cusp boundary is crossed, the lower adiabatic potential no longer has an index-one saddle, so this model no longer contains the associated local saddle-mediated phase-space bottleneck. This is not a statement that electron transfer itself ceases.
\item \textbf{Effect of asymmetry in the model.} At fixed coupling with $0<\delta<1$, increasing $|\eps|/\lr$ tilts the lower sheet until the saddle and one minimum coalesce on a regular branch of the boundary. Beyond the boundary the lower sheet is single-welled and the local activated saddle-crossing picture is absent; the result should not be described as suppression of the actual electron-transfer process.
\item \textbf{Role of reorganization energy.} The dimensionless ratios $2\dd/\lr$ and $\eps/\lr$ show that $\lr$ sets the natural scale for the topology of this lower adiabatic potential.
\item \textbf{Qualitative relation to mixed valence.} The localized/delocalized language of mixed-valence chemistry motivates the model, but no compound-specific Robin--Day assignment follows from the present calculation. The criterion classifies the critical-point structure of the minimal lower adiabatic potential, not mixed-valence materials in general.
\end{enumerate}

Thus, within this finite-dimensional conservative model only, the Marcus--Hush parameters determine whether the lower adiabatic potential supports an index-one saddle and its associated local Hamiltonian transition-state geometry. They do not by themselves establish a solvent free-energy barrier, a rate law, or the validity of Marcus kinetics in a real condensed-phase system.

\section{Local normal form near the saddle-centre equilibrium}

Once the cusp criterion is satisfied, the lower-sheet Hamiltonian $H_-$ possesses a saddle--centre equilibrium $(y_s,0,0,0)$. The local transition-state construction uses the saddle--centre character of this equilibrium. We distinguish the exact separability of the present minimal Marcus Hamiltonian, the quadratic saddle--centre normal form, and the nonlinear terms that remain in the reactive coordinate.

We begin directly from the lower-sheet Hamiltonian introduced in Eq.~\eqref{eq:lower-sheet-Hamiltonian},
\begin{equation}
H_-(y,z,p_y,p_z)
=
\frac{1}{2}p_y^2+
\frac{1}{2}p_z^2+
V_-(y,z).
\label{eq:section5-start-H}
\end{equation}
Using Eq.~\eqref{eq:lower-sheet}, the lower adiabatic potential separates as
\begin{equation}
V_-(y,z)=v_-(y)+\frac{1}{2}\omega^2z^2,
\label{eq:Vminus-separated-vminus}
\end{equation}
where
\begin{equation}
v_-(y)
=
\frac{1}{2}\Omega^2(y^2+y_0^2)
+
\frac{\eps}{2}
-
R(y),
\label{eq:vminus-y-def}
\end{equation}
and
\begin{equation}
R(y)
=
\left[
\left(\Omega^2y_0y-\frac{\eps}{2}\right)^2
+
\dd^2
\right]^{1/2}.
\label{eq:R-y-def}
\end{equation}
Thus $R(y)$ is the square-root splitting term obtained when the two-state diabatic potential matrix is diagonalized. The transverse coordinate remains an exactly harmonic oscillator, while all nonlinearity associated with the lower adiabatic Marcus sheet is contained in the one-dimensional reactive potential $v_-(y)$.

Let $y_s$ denote the saddle point of $v_-(y)$, so that
\begin{equation}
v_-'(y_s)=0,
\qquad
v_-''(y_s)<0,
\label{eq:saddle-vminus-conditions}
\end{equation}
and let
\begin{equation}
E^\ddagger=v_-(y_s)=V_-(y_s,0)
\label{eq:saddle-energy}
\end{equation}
be the saddle energy. Introducing the shifted coordinates
\begin{equation}
Y=y-y_s,
\qquad
Z=z,
\label{eq:translated-coordinates}
\end{equation}
the Hamiltonian can be written exactly as
\begin{equation}
H_-
=
E^\ddagger
+
\frac{1}{2}p_Y^2
+
\bigl[v_-(y_s+Y)-v_-(y_s)\bigr]
+
\frac{1}{2}p_Z^2
+
\frac{1}{2}\omega^2Z^2.
\label{eq:exact-separated-H-vminus}
\end{equation}
Expanding the reactive potential about the saddle gives
\begin{equation}
v_-(y_s+Y)-v_-(y_s)
=
-\frac{1}{2}\Lambda^2Y^2+N_s(Y),
\label{eq:vminus-Taylor-saddle}
\end{equation}
where
\begin{equation}
\Lambda^2=-v_-''(y_s)>0
\label{eq:Lambda-def}
\end{equation}
and
\begin{equation}
N_s(Y)
=
\sum_{k\geq 3}\frac{v_-^{(k)}(y_s)}{k!}Y^k.
\label{eq:reactive-nonlinear-remainder}
\end{equation}
Consequently,
\begin{equation}
H_-
=
E^\ddagger
+
\frac{1}{2}p_Y^2
-
\frac{1}{2}\Lambda^2Y^2
+
N_s(Y)
+
\frac{1}{2}p_Z^2
+
\frac{1}{2}\omega^2Z^2.
\label{eq:exact-separated-expansion}
\end{equation}
The nonlinear terms $N_s(Y)$ generally remain, but they involve only the reactive coordinate. The reactive subsystem is therefore separable from the transverse oscillator without being exactly quadratic. Consequently, this two-degree-of-freedom model contains no bath-induced mode coupling, intermode energy redistribution, chaotic transport, or nonintegrable tube dynamics.

The quadratic part of Eq.~\eqref{eq:exact-separated-expansion} is
\begin{equation}
H_2
=
E^\ddagger
+
\frac{1}{2}p_Y^2
-
\frac{1}{2}\Lambda^2Y^2
+
\frac{1}{2}p_Z^2
+
\frac{1}{2}\omega^2Z^2.
\label{eq:quadratic-H}
\end{equation}
This is the usual saddle--centre quadratic Hamiltonian: the $(Y,p_Y)$ variables describe the hyperbolic electron-transfer direction, while the $(Z,p_Z)$ variables describe the elliptic transverse mode. A linear symplectic change of variables in the hyperbolic plane,
\begin{equation}
q_1=\frac{p_Y+\Lambda Y}{\sqrt{2\Lambda}},
\qquad
p_1=\frac{p_Y-\Lambda Y}{\sqrt{2\Lambda}},
\label{eq:hyperbolic-canonical-coordinates}
\end{equation}
gives
\begin{equation}
\frac{1}{2}p_Y^2-\frac{1}{2}\Lambda^2Y^2
=
\Lambda q_1p_1.
\label{eq:hyperbolic-qp-form}
\end{equation}
Similarly, the standard oscillator scaling
\begin{equation}
q_2=\sqrt{\omega}\,Z,
\qquad
p_2=\frac{p_Z}{\sqrt{\omega}},
\qquad
J_2=\frac{1}{2}(q_2^2+p_2^2),
\label{eq:elliptic-canonical-coordinates}
\end{equation}
brings the bath energy to the form $\omega J_2$. The quadratic normal form is therefore
\begin{equation}
H_2
=
E^\ddagger+
\Lambda q_1p_1+
\omega J_2.
\label{eq:quadratic-normal-form}
\end{equation}
In these variables the full separable Hamiltonian has the local form
\begin{equation}
H_-
=
E^\ddagger+
\Lambda q_1p_1+
\omega J_2
+
N_s\!\left(\frac{q_1-p_1}{\sqrt{2\Lambda}}\right).
\label{eq:separable-normal-form-with-remainder}
\end{equation}
Equation~\eqref{eq:quadratic-normal-form} is the quadratic saddle--centre normal form. Equation~\eqref{eq:separable-normal-form-with-remainder} records the nonlinear correction that remains in the reactive coordinate. The minimal Marcus Hamiltonian is therefore not globally linear, and the quadratic normal form is not exact to all orders; it is exactly separable, with nonlinearities confined to the one-dimensional reactive subsystem.

There is one further local simplification that is specific to a one-degree-of-freedom autonomous hyperbolic subsystem. By a further local canonical normalization of the reactive subsystem, one may write the reactive Hamiltonian in the form
\begin{equation}
H_{\mathrm{react}}
=
E^\ddagger+K(QP),
\qquad
K'(0)=\Lambda,
\label{eq:one-dof-hyperbolic-normal-form}
\end{equation}
where the higher-order resonant terms are generally retained in the function $K$. Thus the linearized dynamics is governed by $\Lambda QP$, but the nonlinear dependence of the hyperbolic motion on energy is not, in general, removed by a symplectic transformation. This distinction is important in the Marcus problem because the square-root term $R(y)$ in Eq.~\eqref{eq:R-y-def} produces nonlinear terms in $v_-(y)$ even though the transverse mode remains exactly separable.

The linearised flow has eigenvalues $\pm\Lambda$ and $\pm i\omega$, characteristic of a saddle--centre equilibrium. Standard local NHIM theory therefore applies. In addition, because the present Hamiltonian is separable, the NHIM, its stable and unstable manifolds, and a local dividing surface can be written explicitly in the original physical variables. This is the point used in the next section.

\section{NHIM, stable and unstable manifolds, and dividing surface}

Whenever the parameters lie inside the cusp region, the equilibrium of the lower-sheet Hamiltonian $H_-$ above the saddle $(y_s,0)$ is of saddle--centre type. The existence of a local phase space transition state then follows from the standard theory of normally hyperbolic invariant manifolds near saddle--centre equilibria \cite{UzerEtAl2002,WaalkensSchubertWiggins2008}. In the present two-degree-of-freedom setting, the NHIM is an unstable periodic orbit. The exact separability of the minimal Marcus Hamiltonian makes this construction especially transparent, but the reason for the construction is the saddle--centre normal hyperbolicity, not the absence of nonlinear reactive terms.

\textbf{The activated complex in phase space.} For the exactly separable Hamiltonian, the periodic orbit written below is an exact invariant object for each $E>E^\ddagger$; it is not merely the orbit of the quadratic normal form. We nevertheless use it as a \emph{local} transition-state object in a neighborhood of the saddle, where the saddle-mediated bottleneck interpretation and the adiabatic lower-sheet approximation are intended to apply. The NHIM at energy $E$ is

\begin{equation}
\mathcal{N}_E
=
\left\{
Y=0,
\quad
p_Y=0,
\quad
\frac{1}{2}p_Z^2+
\frac{1}{2}\omega^2Z^2=E-E^\ddagger
\right\}.
\label{eq:exact-nhim-physical-variables}
\end{equation}
Equivalently, in the quadratic normal-form variables this is $q_1=p_1=0$ with $\omega J_2=E-E^\ddagger$. Since the remaining motion is the transverse harmonic oscillator, Eq.~\eqref{eq:exact-nhim-physical-variables} is a periodic orbit. It is unstable because perturbations in the $(Y,p_Y)$ directions grow or decay exponentially at the hyperbolic rate determined by the saddle.

Geometrically, $\mathcal{N}_E$ is the phase space object sitting over the lower-sheet saddle that plays the dynamical role of the activated complex. Trajectories initialized exactly on it remain at the saddle in the electron-transfer coordinate and oscillate in the transverse mode. They neither pass from one well to the other nor leave along a reactive channel.

\textbf{Stable and unstable manifolds.} The stable and unstable manifolds of $\mathcal{N}_E$ organize trajectories approaching and leaving the bottleneck. In the quadratic normal form their leading-order expressions are
\begin{equation}
W^s(\mathcal{N}_E):\ q_1=0,
\qquad
W^u(\mathcal{N}_E):\ p_1=0,
\label{eq:stable-unstable-leading-order}
\end{equation}
with $\omega J_2=E-E^\ddagger$. For the exact separable Marcus Hamiltonian, these manifolds can also be viewed as the product of the zero-hyperbolic-energy stable or unstable separatrix of the one-dimensional reactive subsystem with the transverse oscillator orbit at energy $E-E^\ddagger$. In the original variables, the hyperbolic part is governed by
\begin{equation}
\frac{1}{2}p_Y^2+v_-(y_s+Y)-v_-(y_s)=0.
\label{eq:zero-hyperbolic-energy}
\end{equation}
Thus the nonlinear terms in the reactive coordinate deform the separatrix branches away from their purely linear forms, but they do not destroy the invariant-manifold structure. In a nonseparable extension, the corresponding manifolds would generally require numerical computation and would no longer have this product form.

\textbf{The locally no-recrossing dividing surface.} A local dividing surface attached to the NHIM may be taken, in the physical coordinates, as
\begin{equation}
\DS_E
=
\left\{
Y=0,
\quad
\frac{1}{2}p_Y^2+
\frac{1}{2}p_Z^2+
\frac{1}{2}\omega^2Z^2=E-E^\ddagger
\right\}.
\label{eq:exact-dividing-surface}
\end{equation}
The full dividing surface $\DS_E$ is a closed two-sphere and has no boundary. The NHIM \eqref{eq:exact-nhim-physical-variables}, obtained by setting $p_Y=0$, is its equator and is the common boundary of the closures of its two hemispheres. Their interiors are distinguished by the sign of $p_Y$:
\begin{equation}
\DS_E^+ = \DS_E\cap\{p_Y>0\},
\qquad
\DS_E^- = \DS_E\cap\{p_Y<0\}.
\label{eq:forward-backward-hemispheres}
\end{equation}
Since $\dot{Y}=p_Y$, the vector field is strictly transverse to the interior of the two hemispheres, where $p_Y\neq0$, and crosses them with opposite orientations. In the present separable model, both Eq.~\eqref{eq:exact-dividing-surface} and this transversality statement are exact rather than quadratic-normal-form approximations. The no-recrossing claim is local to a passage through the bottleneck: trajectories may return to the same surface after a global excursion. We make no claim here that this surface is a globally no-recrossing or globally minimal-flux surface over the full energy shell.

\textbf{Consequences for the lower-sheet model.} For parameters satisfying Eq.~\eqref{eq:cusp-criterion-dimensional}, the lower-sheet Hamiltonian has a saddle--centre equilibrium. Exact separability gives the periodic orbit $\mathcal{N}_E$, its stable and unstable manifolds, and the attached dividing surface for each $E>E^\ddagger$. The construction describes local passage through the neighborhood of the saddle and does not exclude later global returns. Its physical interpretation is also restricted to energies and parameter regimes for which classical adiabatic propagation on the lower sheet remains appropriate.

Inside the cusp region, the lower adiabatic Marcus surface therefore does not merely possess an energetic barrier in the reduced-coordinate sense; it supports a local phase space bottleneck. Outside the cusp, the lower sheet lacks the index-one saddle needed to anchor this local NHIM-based transition-state structure.

\section{Illustrative visualization of the analytic cusp geometry}

\begin{figure}[htbp]
\centering
\includegraphics[width=0.82\textwidth]{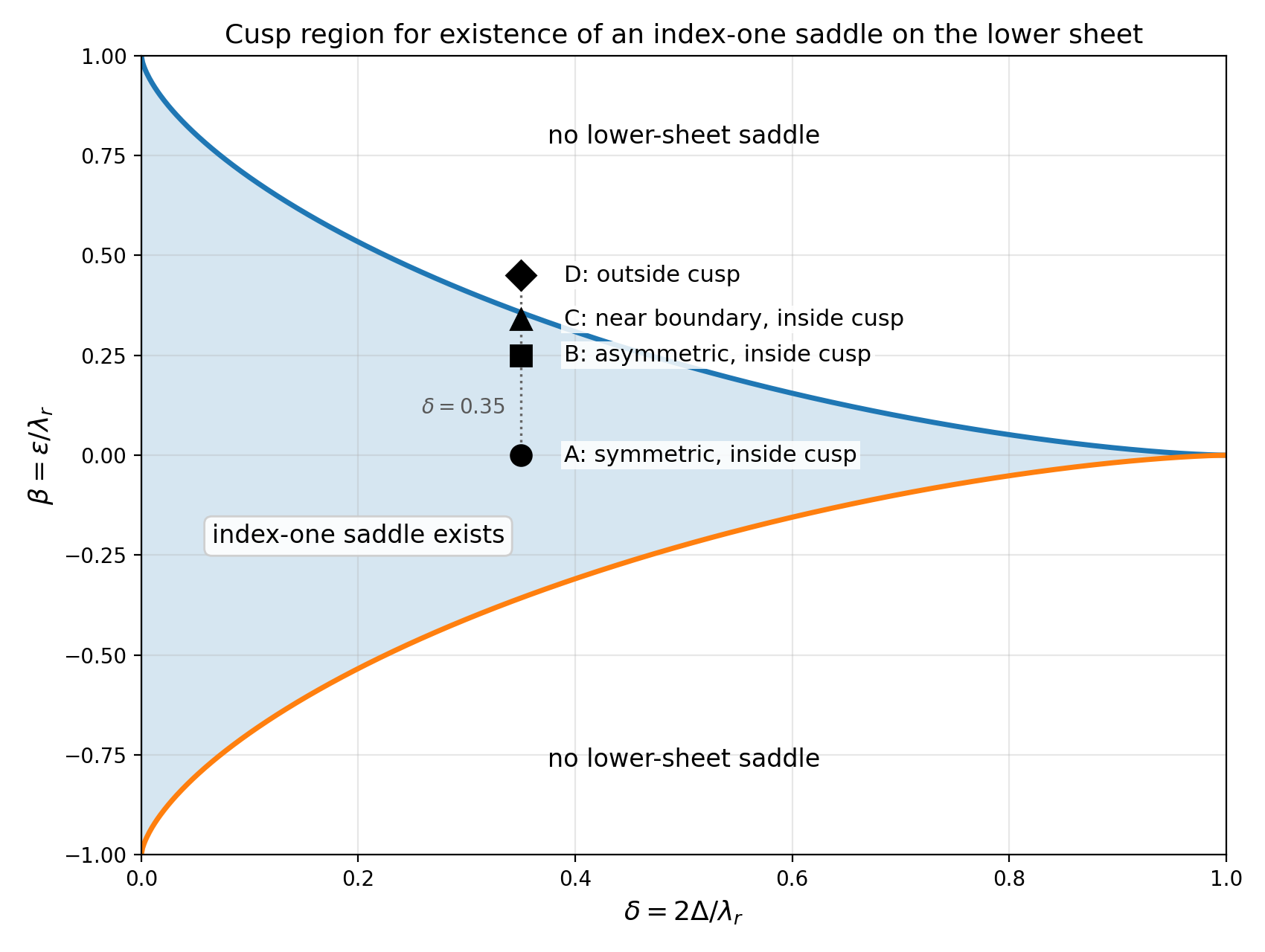}
\caption{Cusp region in the $(\beta,\delta)$-plane for the existence of an index-one saddle on the lower adiabatic sheet. The shaded region satisfies $|\beta|<(1-\delta^{2/3})^{3/2}$ with $0<\delta<1$ and therefore supports a local Hamiltonian bottleneck. Outside the cusp there is no lower-sheet saddle and hence no such local bottleneck. The marked parameter values, all on the line $\delta=0.35$, are A: $(\beta,\delta)=(0,0.35)$, symmetric inside cusp; B: $(0.25,0.35)$, asymmetric inside cusp; C: $(0.34,0.35)$, near the cusp boundary but still inside; and D: $(0.45,0.35)$, outside cusp. These same cases are used in Figure~\ref{fig:lower-sheet}.}
\label{fig:cusp}
\end{figure}

\begin{figure}[htbp]
\centering
\includegraphics[width=\textwidth]{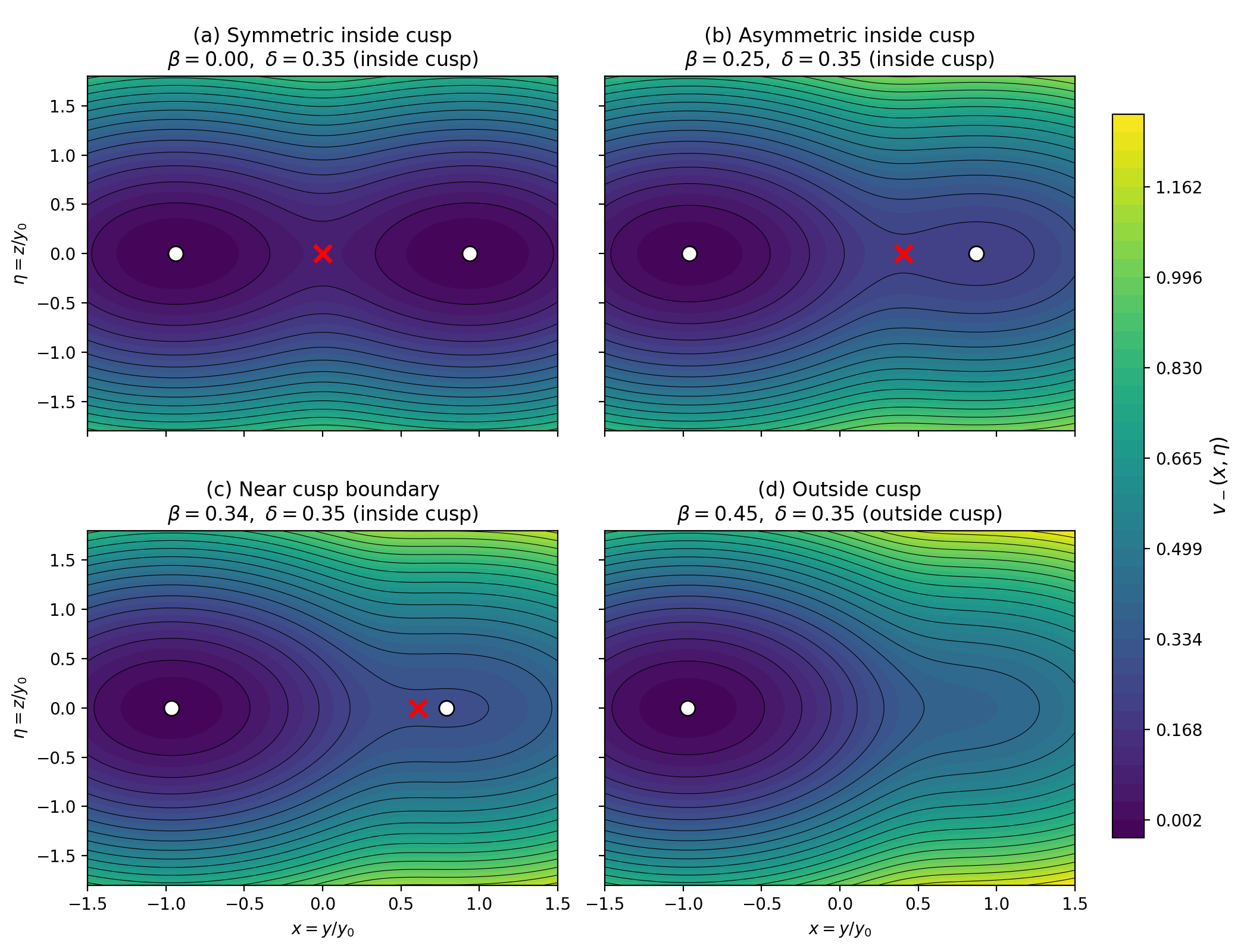}
\caption{Contours of the dimensionless lower adiabatic sheet for the four parameter values A--D marked in Figure~\ref{fig:cusp}. Open circles denote minima and red crosses denote saddles. Panels (a)--(c) lie inside the cusp and therefore exhibit a local lower-sheet bottleneck; panel (d) lies outside the cusp and contains only a single minimum. The figure visualizes how asymmetry shifts the saddle and how the bottleneck is lost as the cusp boundary is approached and crossed.}
\label{fig:lower-sheet}
\end{figure}

\begin{figure}[htbp]
\centering
\includegraphics[width=\textwidth]{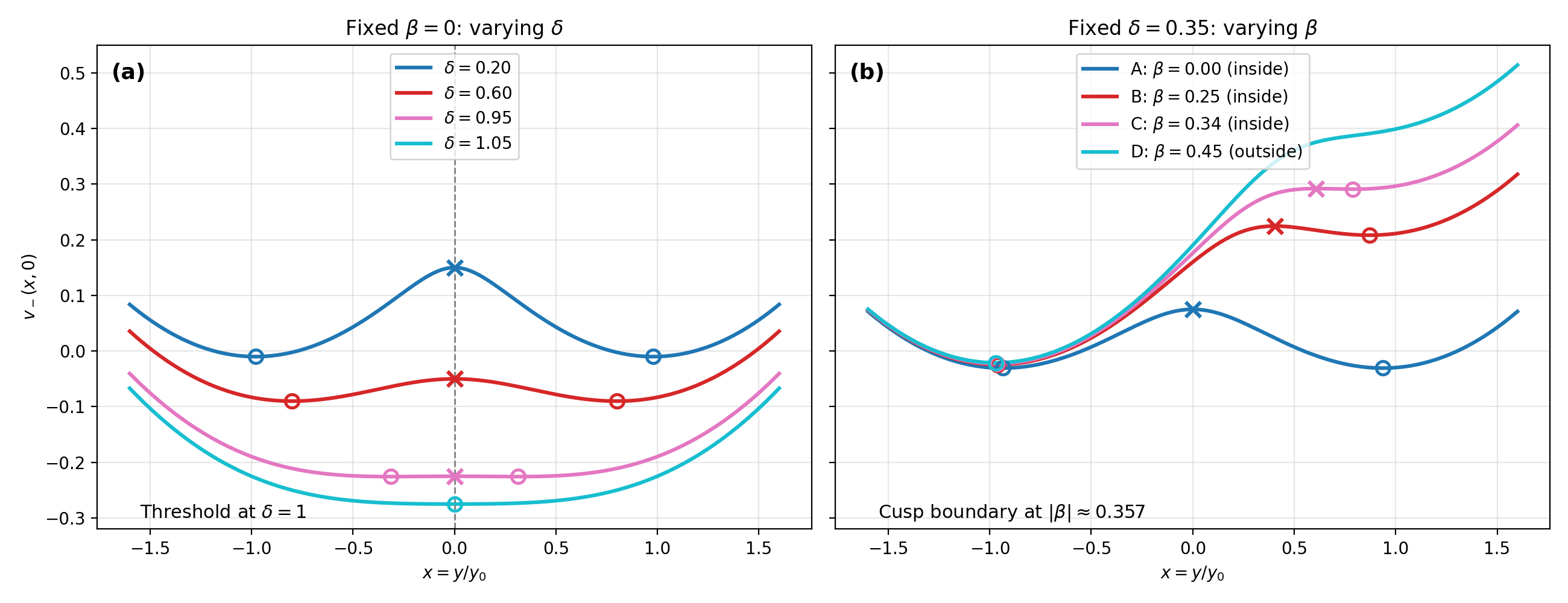}
\caption{One-dimensional cuts of the lower adiabatic sheet $v_-(x,0)$. Open circles denote minima and crosses denote saddles. Panel (a) shows the symmetric case $\beta=0$ for $\delta=0.20$, $0.60$, $0.95$, and $1.05$, straddling the threshold $\delta=1$. Panel (b) fixes $\delta=0.35$ and uses the same A--D asymmetry values marked in Figure~\ref{fig:cusp}. The cuts show the changes in stationary-point structure across the cusp boundary.}
\label{fig:cuts}
\end{figure}

The calculations in this section evaluate and display the analytic formulas derived above. Figures~\ref{fig:cusp}--\ref{fig:cuts} directly illustrate the cusp boundary and the associated lower-sheet critical-point structure. They are not presented as independent numerical validation. The purely illustrative phase-space material---the NHIM/UPO, separatrix, dividing-surface, and short-trajectory visualizations---is collected in Appendix~\ref{app:additional-geometry}. No reactive-flux estimate, recrossing statistic, gap-time distribution, or compound-specific rate calculation is inferred from these figures.

The dimensionless lower-sheet Hamiltonian used in the figures is:
\begin{equation}
\mathcal{H}(x,\eta,p_x,p_\eta)
=
\frac{1}{2} p_x^2+\frac{1}{2} p_\eta^2+u(x)+\frac{\nu^2}{4}\eta^2,
\qquad
\nu=\frac{\omega}{\Omega},
\label{eq:numerical-H}
\end{equation}
where the reduced lower-sheet potential $u(x)$ is given by:
\begin{equation}
 u(x)=\frac14(x^2+1)+\frac{\beta}{2}-\frac{1}{2}\sqrt{(x-\beta)^2+\delta^2},
\label{eq:numerical-u}
\end{equation}
using the dimensionless parameters $x=y/y_0$, $\eta=z/y_0$, $\beta=\eps/\lr$, and $\delta=2\dd/\lr$. In all following figures, we fix the frequency ratio $\nu=1$ so that the transverse oscillator scale is comparable to the electron-transfer scale. 

\textbf{Cusp geometry and the shape of the lower sheet.} Figure~\ref{fig:cusp} plots the cusp boundary ($|\beta|=(1-\delta^{2/3})^{3/2}$) and shades the region in which the lower adiabatic sheet has an index-one saddle. The marked parameter values A--D are used in Figure~\ref{fig:lower-sheet} and in panel (b) of Figure~\ref{fig:cuts}; all four have $\delta=0.35$, with the asymmetry varied through the values stated in the caption.

Figure~\ref{fig:lower-sheet} translates this parameter-space criterion into the actual topography of the lower sheet. The four panels correspond to the four points A--D in Figure~\ref{fig:cusp}: a symmetric barrier-bearing case, an asymmetric barrier-bearing case, a near-cusp interior case where the saddle and one minimum are close to coalescing, and an outside-cusp case containing only a single minimum. 

Figure~\ref{fig:cuts} shows one-dimensional cuts of the lower sheet at $\eta=0$.

These one-dimensional cuts give the reduced-coordinate potential profiles used in chemistry \cite{Nitzan2006}. They show how asymmetry or coupling removes the barrier. The phase space analysis identifies the associated change in the local saddle--centre and transition-state geometry.

\FloatBarrier

\section{Discussion and outlook}

We have considered a finite-dimensional conservative model with classical adiabatic motion on the lower sheet. The local phase space transition-state construction applies near an index-one saddle \cite{WaalkensSchubertWiggins2008}; the question is when the lower adiabatic Marcus surface possesses such a saddle. For the minimal asymmetric two-degree-of-freedom model, the answer is given by the analytic cusp criterion \eqref{eq:cusp-criterion-dimensional}.

This distinction clarifies what is standard and what is added here. The lower-sheet double-well disappearance boundary is known in electron-transfer theory \cite{JungCao2002,Nitzan2006,MayKuhn2011}, and the NHIM/dividing-surface construction near a saddle--centre equilibrium is established reaction-dynamics theory \cite{UzerEtAl2002,WaalkensSchubertWiggins2008}. The contribution of the present paper is the direct admissible-root and stability analysis showing that, in this minimal conservative Hamiltonian, the known lower-sheet boundary is also exactly the boundary for the existence of the local saddle--centre/NHIM transition-state geometry.

This statement should not be conflated with the usual Marcus free-energy picture. The present $V_-$ is a finite-dimensional potential-energy surface, not a solvent free-energy surface or potential of mean force. Outside the cusp, this model lacks a lower-sheet index-one saddle and the corresponding local Hamiltonian transition-state structure; no conclusion follows about the existence or rate of electron transfer in a dissipative condensed-phase system.

\textbf{Relation to Hamiltonian saddle-node geometry.} The cusp boundary also suggests a useful connection with the Hamiltonian saddle-node analysis of Garc\'ia-Garrido, Naik, and Wiggins \cite{GarciaGarridoNaikWiggins2020}. In that work, the emphasis was not only on visualizing the NHIM, stable and unstable manifolds, and dividing surfaces, but also on how those structures change as a saddle and a centre coalesce. At regular points of its cusp boundary, the present Marcus model contains the same local bifurcation mechanism, but with the bifurcation parameter determined by the Marcus--Hush quantities $\eps$, $\dd$, and $\lr$.

Near a regular point of the cusp curve in the $(\beta,\delta)$-plane, the lower-sheet reactive potential can be reduced locally to a one-degree-of-freedom saddle-node normal form
\begin{equation}
H_{\mathrm{SN}}
=
\frac{1}{2}P^2+a\sigma Q+\frac{b}{3}Q^3+\cdots,
\label{eq:discussion-saddle-node-normal-form}
\end{equation}
where $\sigma$ is a signed distance from the cusp boundary, chosen positive on the three-root side, and $a$ and $b$ are nonzero model-dependent coefficients. Near a regular boundary point, this normal form implies the usual saddle-node scalings, up to model-dependent constants: the saddle-minimum separation scales as $\sigma^{1/2}$, the barrier height scales as $\sigma^{3/2}$, and the hyperbolic exponent at the saddle scales as $\sigma^{1/4}$. In the Marcus setting these scalings would quantify how the activated lower-sheet bottleneck collapses as electronic coupling or asymmetry drives the model toward the cusp boundary.

The separable form of the minimal Marcus Hamiltonian makes this analysis considerably easier than the more general coupled saddle-node model. The NHIM/UPO, its invariant manifolds, and the local dividing surface can be obtained from explicit formulas once $y_s$ and $E^\ddagger$ are known; differential correction, continuation, and manifold globalization are not required for the present two-degree-of-freedom separable model. Because the transverse mode is exactly separable, there is no exchange of energy between the bath coordinate and the electron-transfer coordinate, and therefore no nonintegrable tube dynamics. The ``tilting and squeezing'' mechanisms studied in Ref.~\cite{GarciaGarridoNaikWiggins2020} would become chemically more significant in a nonseparable Marcus or system-bath extension, for example with unequal Hessians, Duschinsky rotation, anharmonic solvent modes, or explicit coupling between the energy-gap coordinate and additional bath coordinates.

For this reason, the near-cusp saddle-node analysis is best viewed as a future direction rather than as part of the claim established here. The present result is the Hamiltonian phase-space interpretation and direct root/stability classification of the known lower-sheet cusp boundary in the minimal separable model. Explicit scaling laws for flux, reactive fractions, or coupled bath dynamics would require additional analysis and are not claimed in this paper.

\textbf{Limitations and future directions.} Extending the present Hamiltonian treatment to solvent dynamics and nonadiabatic transfer raises three distinct problems:

\begin{enumerate}
    \item \textbf{Dimensionality Reduction:} We have not attempted here to reduce the many-solvent problem to an energy-gap coordinate or potential of mean force. While such dimensionality reductions are central to practical simulations \cite{NewtonSutin1984}, they intentionally integrate out the transverse phase space structures (like the NHIM) that we aimed to expose here.
    \item \textbf{Dissipative Open Systems:} Dissipative solvent theories, such as the Zusman or Sumi--Marcus models \cite{Zusman1980,SumiMarcus1986}, and quantum system--bath descriptions, such as the spin-boson treatment of the dissipative two-state system \cite{Garg1985,Leggett1987}, include environmental dynamics or their reduced influence. Their physical assumptions and dynamical variables differ from those of the present classical, conservative two-degree-of-freedom lower-sheet Hamiltonian.
    \item \textbf{Nonadiabatic Dynamics:} Nonadiabatic electron transfer requires a treatment that allows transitions between electronic states. Examples include the quantum-classical Liouville equation (QCLE), continuous mapping approaches such as MMST, and surface-hopping approaches such as MASH \cite{Kapral2006,LawrenceMannouchRichardson2024}. These effects are excluded by the restriction to classical propagation on the lower adiabatic sheet.
\end{enumerate}

These distinctions point directly to the necessary next steps. The most immediate continuation is to move from the present two-degree-of-freedom model to a finite-dimensional system-bath Hamiltonian with multiple transverse modes. This would produce higher-dimensional NHIMs and allow one to study how complex solvent environments deform the bottleneck and induce recrossings away from the strict local regime. A subsequent conceptual step is to investigate how the lower-sheet bottleneck criterion formulated here changes once dissipative friction or nonadiabatic surface-hopping effects are reintroduced.

We do not compute rates, gap-time distributions, global reactant volumes, bath-induced energy redistribution, or compound-specific mixed-valence observables. For the minimal conservative lower-sheet model, we have established that the known cusp boundary for loss of the double well is also the boundary for the existence of an index-one saddle and the associated local Hamiltonian NHIM/dividing-surface geometry.

\clearpage
\appendix

\section{Direct admissible-root count}
\label{app:root-count}

This appendix supplies a direct root count that avoids any reliance on the quartic roots introduced by squaring. Define
\begin{equation}
F(x;\beta,\delta)
=
x-\frac{x-\beta}{\sqrt{(x-\beta)^2+\delta^2}}.
\label{eq:appendix-direct-root-function}
\end{equation}
The critical points of the lower sheet are exactly the zeros of $F$. Differentiation gives
\begin{equation}
F'(x)
=
1-\frac{\delta^2}{\bigl((x-\beta)^2+\delta^2\bigr)^{3/2}}.
\label{eq:appendix-direct-root-derivative}
\end{equation}
If $\delta\geq1$, then $F'(x)\geq0$ for all $x$ (strictly positive except at one isolated point when $\delta=1$). Since $F(x)\to-\infty$ as $x\to-\infty$ and $F(x)\to+\infty$ as $x\to+\infty$, $F$ has exactly one zero. This zero is simple except at $(\beta,\delta)=(0,1)$, where $F(x)=x^3/2+O(x^5)$ and $x=0$ is a triple root. At this cusp tip the potential has a degenerate minimum, with $u(x)=-1/4+x^4/16+O(x^6)$, rather than a saddle--centre equilibrium.

Now suppose $0<\delta<1$. Equation~\eqref{eq:appendix-direct-root-derivative} vanishes at exactly two points,
\begin{equation}
x_\pm
=
\beta\pm\delta^{2/3}\sqrt{1-\delta^{2/3}},
\label{eq:appendix-direct-root-extrema}
\end{equation}
where $x_-$ is a local maximum and $x_+$ is a local minimum. Direct substitution yields
\begin{equation}
F(x_-)=\beta+\bigl(1-\delta^{2/3}\bigr)^{3/2},
\qquad
F(x_+)=\beta-\bigl(1-\delta^{2/3}\bigr)^{3/2}.
\label{eq:appendix-direct-root-extrema-values}
\end{equation}
Because $F$ is increasing on $(-\infty,x_-)$, decreasing on $(x_-,x_+)$, and increasing on $(x_+,\infty)$, it has exactly three zeros precisely when
\begin{equation}
F(x_-)>0>F(x_+),
\end{equation}
which is equivalent to
\begin{equation}
|\beta|<\bigl(1-\delta^{2/3}\bigr)^{3/2}.
\end{equation}
For $0<\delta<1$, equality places one extremum on the axis, giving one simple root and one double root; outside this interval there is exactly one simple root. Finally, since the reduced potential satisfies $u'(x)=F(x)/2$ and $u''(x)=F'(x)/2$, the three roots inside the cusp are ordered minimum--saddle--minimum. This proves necessity and sufficiency over the full parameter range and excludes extraneous roots without invoking the squared quartic equation.

\section{Additional illustrative phase-space geometry}
\label{app:additional-geometry}

The following four figures and their associated discussion are supplementary illustrations of the exact separable formulas. They are not presented as trajectory-based validation or as calculations of reactive flux, recrossing statistics, or gap times.

\begin{figure}[htbp]
\centering
\includegraphics[width=\textwidth]{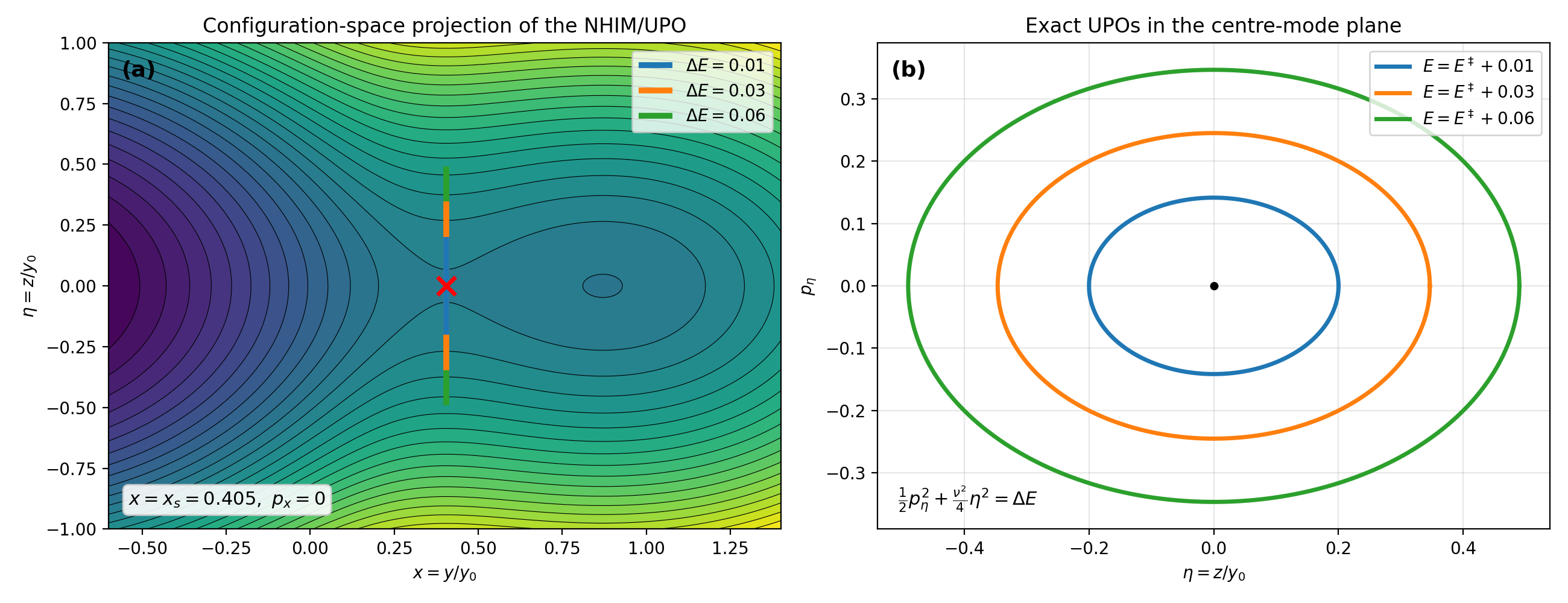}
\caption{NHIM/UPO above the lower-sheet saddle for the representative inside-cusp parameter set $(\beta,\delta,\nu)=(0.25,0.35,1)$. Panel (a) shows the configuration-space projection of the NHIM/UPO for three energies $E=E^\ddagger+\Delta E$ with $\Delta E=0.01$, $0.03$, and $0.06$, showing that in the minimal separable model the orbit lies at fixed $x=x_s$ over the saddle. Panel (b) shows the same UPOs in the center-mode plane $(\eta,p_\eta)$, where they form exact ellipses determined by $\frac{1}{2} p_\eta^2+\frac{\nu^2}{4}\eta^2=\Delta E$. The simplicity of the orbit reflects the separability of the minimal lower-sheet Hamiltonian.}
\label{fig:nhim-upo}
\end{figure}

\begin{figure}[htbp]
\centering
\includegraphics[width=\textwidth]{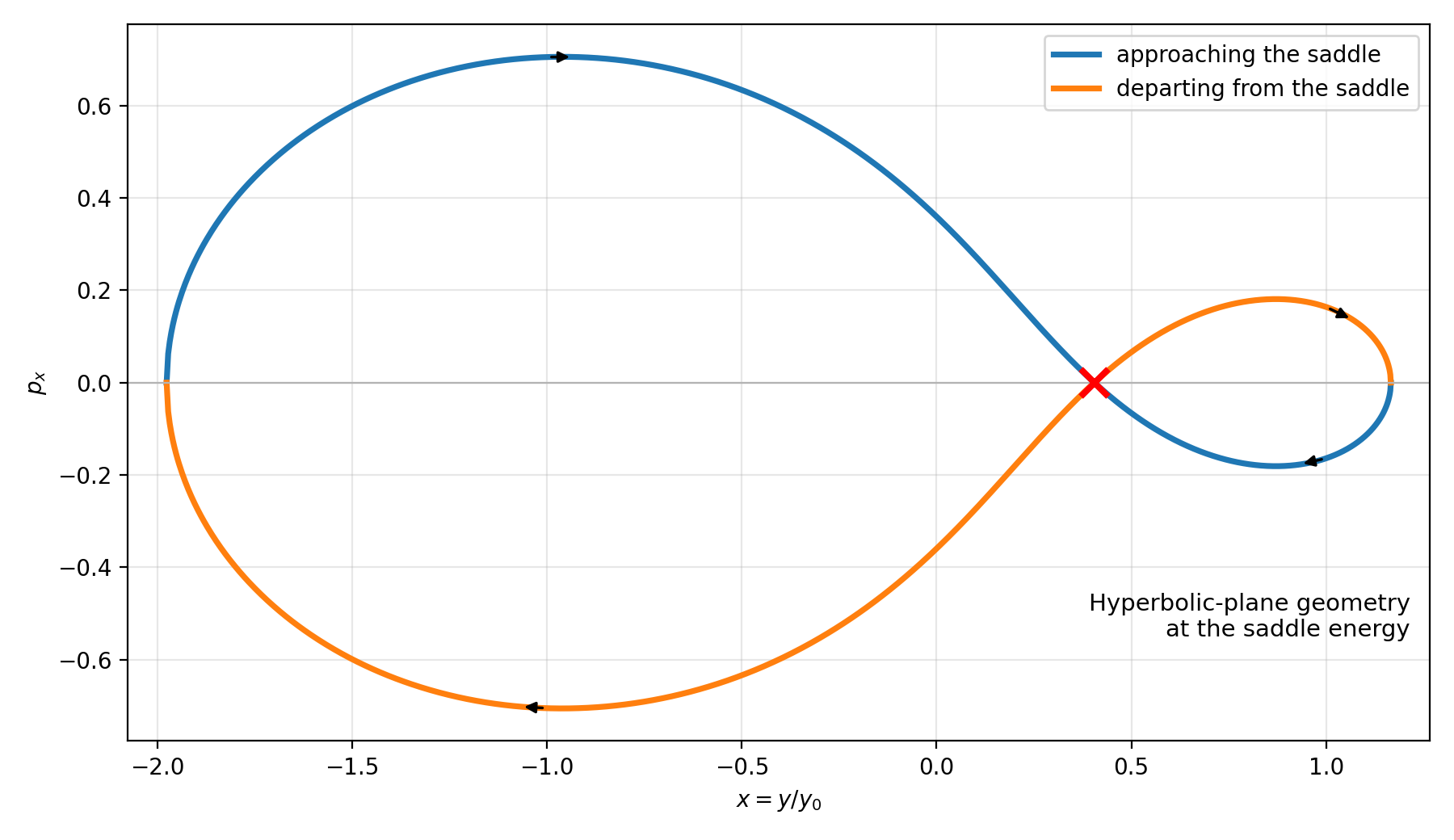}
\caption{Zero-reactive-energy separatrix for the representative inside-cusp parameter set $(\beta,\delta,\nu)=(0.25,0.35,1)$. The curve is shown in the hyperbolic $(x,p_x)$ plane at the saddle energy and is determined by $p_x=\pm[2(E^\ddagger-u(x))]^{1/2}$. In the minimal separable model the global stable and unstable manifolds coincide as sets in this projection. Blue denotes portions moving toward the saddle and orange portions moving away from it, with arrows indicating the direction of motion. These are the local stable and unstable branches near the saddle, but each complete homoclinic loop belongs to both global manifolds. The horizontal scale shows the complete figure-eight: the larger loop around the left minimum, out to the turning point $x_L=-1.976$, and the smaller loop around the right minimum, out to $x_R=1.166$, where $u(x_L)=u(x_R)=E^\ddagger$.}
\label{fig:manifolds}
\end{figure}

\begin{figure}[htbp]
\centering
\includegraphics[width=\textwidth]{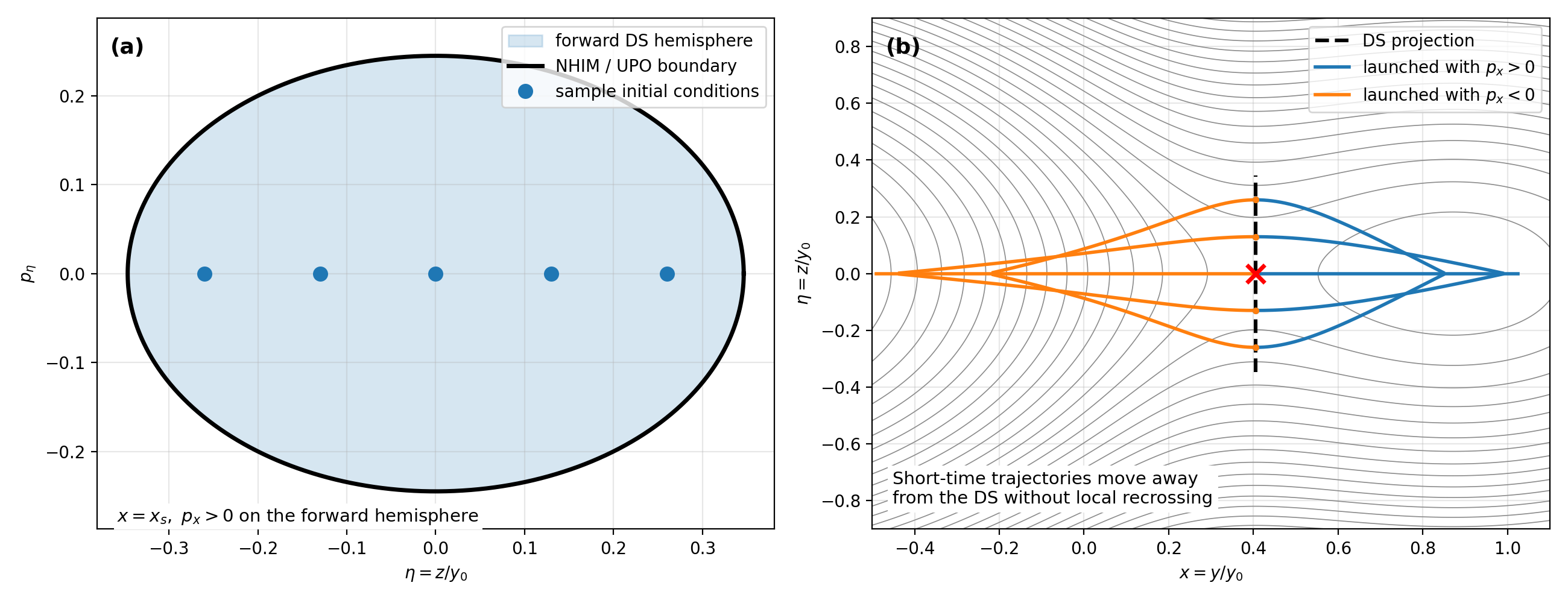}
\caption{Dividing surface attached to the NHIM/UPO for the representative inside-cusp parameter set at energy $E=E^\ddagger+0.03$. Panel (a) shows a dividing-surface hemisphere represented in the $(\eta,p_\eta)$ plane at $x=x_s$, with the NHIM/UPO as its elliptical boundary. Interior sample points have $p_x\ne 0$ and are not on the UPO. Panel (b) shows short-time trajectories launched from points on the dashed vertical line $x=x_s$, projected into configuration space $(x,\eta)$. The initial points are marked on the line, and the trajectories move away from it without immediate local recrossing.}
\label{fig:ds}
\end{figure}

\begin{figure}[htbp]
\centering
\includegraphics[width=\textwidth]{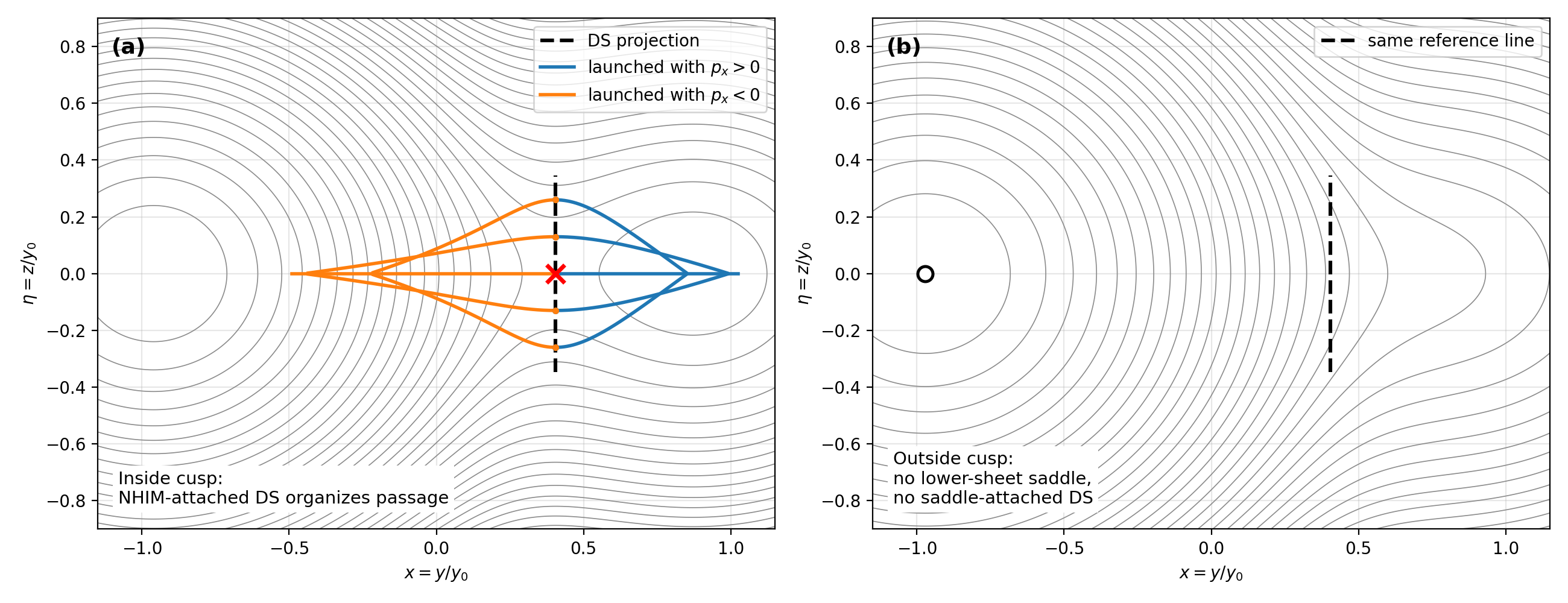}
\caption{Inside-cusp versus outside-cusp comparison at fixed $\delta=0.35$. Panel (a) uses the inside-cusp parameter set $(\beta,\delta)=(0.25,0.35)$, for which the dashed line is the projection of a dividing surface attached to the NHIM over the lower-sheet saddle. Panel (b) uses the outside-cusp parameter set $(\beta,\delta)=(0.45,0.35)$; the same vertical line is now only a reference line and not a transition-state object. Panel (b) has no saddle-attached local transition-state structure.}
\label{fig:inside-outside}
\end{figure}

\textbf{Transition-state geometry in the minimal model.} In the dimensionless formulation \eqref{eq:numerical-H}, the lower-sheet Hamiltonian is exactly separable. For an inside-cusp parameter set, the NHIM (an unstable periodic orbit, or UPO, in two degrees of freedom) follows analytically from the saddle location, without numerical shooting or differential correction. For Figures~\ref{fig:nhim-upo}--\ref{fig:inside-outside}, we use the representative inside-cusp parameter set $(\beta,\delta,\nu)=(0.25,0.35,1)$, which places the saddle at $x_s\approx 0.405$.

\begin{samepage}
Figure~\ref{fig:nhim-upo} shows the resulting NHIM/UPO for three energies above the barrier ($E=E^\ddagger+\Delta E$). Because the model is separable, the configuration-space projection of the orbit sits at a fixed $x=x_s$ over the saddle, growing only in the transverse direction. In the center-mode plane $(\eta,p_\eta)$, these orbits are exact ellipses. 

\end{samepage}

Figure~\ref{fig:manifolds} shows the zero-reactive-energy separatrix branches through the saddle in the hyperbolic $(x,p_x)$ plane. For this integrable separable model the global stable and unstable manifolds coincide as sets in this projection: each complete homoclinic loop approaches the saddle in both time directions. The colors distinguish the portions instantaneously approaching or departing from the saddle; near the saddle these are the local stable and unstable branches, respectively. These exact separatrix curves make a separate Poincar\'e-map construction unnecessary here.

The visual simplicity of these structures is a direct consequence of exact separability. The figures illustrate the analytic invariant objects of this integrable benchmark; they do not demonstrate mode coupling, energy redistribution, manifold splitting, or nonintegrable transport. Such phenomena require coupling between the reactive and transverse modes and are outside the scope of the present paper.

\FloatBarrier

\textbf{Visualizing the locally no-recrossing dividing surface.} Figure~\ref{fig:ds} shows the dividing surface attached to the NHIM/UPO. Panel (a) represents one dividing-surface hemisphere; its elliptical boundary is precisely the NHIM, where $p_x=0$. The sample points in the interior of the ellipse have $p_x\ne 0$ and are therefore not on the UPO. Panel (b) launches short-time trajectories from these interior dividing-surface points on the dashed vertical line $x=x_s$. The trajectories move away from the dividing surface on opposite sides according to the sign of $p_x$, without immediate local recrossing.

Figure~\ref{fig:inside-outside} compares an inside-cusp parameter set with an outside-cusp parameter set at the same coupling strength ($\delta=0.35$). Inside the cusp (panel a), the dashed line is the projection of an NHIM-attached dividing surface, and trajectories launched from the two hemispheres move to opposite sides of the local bottleneck. Outside the cusp (panel b), the same vertical line is only a reference line. There is no lower-sheet saddle, no NHIM above it, and no analogous local transition-state structure. The cusp criterion therefore determines whether this local phase space transport structure exists.

\FloatBarrier

\section{Dimensionless numerical formulation and reproducibility}
\label{app:numerics}

This appendix records the formulas, parameter choices, and numerical tolerances used to construct the three main-text cusp/potential illustrations and the four additional phase-space illustrations collected in Appendix~\ref{app:additional-geometry}. The calculations reproduce the analytic figures and short-time trajectories; they do not validate transport rates or long-time recrossing behavior. The scripts that generate every figure, together with a diagnostics script that prints the quantities quoted below, are provided as described in the Data and code availability statement.

\textbf{Dimensionless model and equations of motion.} For numerical work, we scale the electron-transfer and transverse coordinates by $y_0$ and use the dimensionless asymmetry and coupling parameters:
\[
x=\frac{y}{y_0},\qquad \eta=\frac{z}{y_0},\qquad \beta=\frac{\eps}{\lr},\qquad \delta=\frac{2\dd}{\lr},\qquad \nu=\frac{\omega}{\Omega}.
\]
Specifically, with $t_{\mathrm{phys}}$ denoting dimensional time, we use
\[
p_x=\frac{p_y}{\sqrt{\lr}},\qquad
p_\eta=\frac{p_z}{\sqrt{\lr}},\qquad
t=\frac{\sqrt{\lr}}{y_0}\,t_{\mathrm{phys}},\qquad
\mathcal{H}=\frac{H_-}{\lr}.
\]
The lower-sheet dynamics then takes the dimensionless form:
\begin{equation}
\mathcal{H}(x,\eta,p_x,p_\eta)=\frac{1}{2} p_x^2+\frac{1}{2} p_\eta^2+u(x)+\frac{\nu^2}{4}\eta^2,
\label{eq:appendix-H}
\end{equation}
where the reduced 1D potential $u(x)$ is:
\begin{equation}
 u(x)=\frac14(x^2+1)+\frac{\beta}{2}-\frac{1}{2}\sqrt{(x-\beta)^2+\delta^2}.
\label{eq:appendix-u}
\end{equation}
Hamilton's equations are:
\begin{equation}
\dot{x}=p_x,
\qquad
\dot{\eta}=p_\eta,
\qquad
\dot{p}_x=-u'(x),
\qquad
\dot{p}_\eta=-\frac{\nu^2}{2}\eta,
\label{eq:appendix-eom}
\end{equation}
where the force along the electron-transfer coordinate relies on the derivatives:
\begin{equation}
 u'(x)=\frac{1}{2} x-\frac{1}{2}\frac{x-\beta}{\sqrt{(x-\beta)^2+\delta^2}},
\qquad
 u''(x)=\frac{1}{2}-\frac{1}{2}\frac{\delta^2}{\bigl((x-\beta)^2+\delta^2\bigr)^{3/2}}.
\label{eq:appendix-u-primes}
\end{equation}

\textbf{Locating critical points and the saddle.} For any fixed parameter pair $(\beta,\delta)$, the critical points satisfy $u'(x)=0$, so every critical point is among the roots of the squared quartic equation below; the candidates must also satisfy the original unsquared equation:
\begin{equation}
x^4-2\beta x^3+(\beta^2+\delta^2-1)x^2+2\beta x-\beta^2=0.
\label{eq:appendix-quartic}
\end{equation}
Our numerical routines solve Eq.~\eqref{eq:appendix-quartic} with the companion-matrix polynomial root finder of \texttt{NumPy} (\texttt{numpy.roots}). Because squaring introduces roots that do not satisfy the original critical-point equation, each of the four candidates is filtered as follows: (i) candidates with $|\operatorname{Im}x|>10^{-8}$ are discarded; (ii) the real part of each remaining candidate is accepted only if $|u'(x)|\le10^{-10}$, equivalently $|F(x;\beta,\delta)|\le2\times10^{-10}$ for the original unsquared equation \eqref{eq:critical-equation-dimensionless}, which removes extraneous real roots created by squaring; (iii) a candidate separated by less than $10^{-8}$ from an already accepted root is merged with it. For the symmetric figure parameters ($\beta=0$, $\delta\ne1$), the quartic factorizes as $x^2(x^2+\delta^2-1)$, and its double root at $x=0$ represents one simple root of $u'(x)=0$. The cusp tip $(\beta,\delta)=(0,1)$ is excluded from the plotted parameter sets; there the quartic has a quadruple root at $x=0$ and the unsquared equation has a triple root. Accepted roots are classified as minima or saddles by the sign of $u''(x)$. For all parameter sets used in the figures, the accepted roots satisfy $|u'(x)|<10^{-14}$ in the tested software environments, while every real candidate rejected as extraneous has $|u'(x)|>0.18$; duplicate candidates are counted separately and can have zero residual. The smallest $|u''(x)|$ at an accepted root is approximately $0.024$ (for the near-threshold symmetric cases of Figure~\ref{fig:cuts}(a); it is $0.167$ for the cases A--D). These margins make the plotted root counts and classifications insensitive to small changes in the stated tolerances. This numerical check concerns the specified figure parameters; degenerate roots on the bifurcation boundary are classified analytically in Appendix~\ref{app:root-count}. The candidate-by-candidate table for all parameter sets is written by the diagnostics script of the accompanying code. 

For the phase space visualizations (Figures~\ref{fig:nhim-upo}--\ref{fig:inside-outside}), we use the representative inside-cusp parameter set:
\begin{equation}
(\beta,\delta,\nu)=(0.25,0.35,1).
\label{eq:appendix-main-params}
\end{equation}
For these parameters the procedure above returns three critical points, the minima $x=-0.960661$ ($u=-0.024402$) and $x=0.871246$ ($u=0.208240$) and the saddle
\begin{equation}
x_s=0.405062,
\qquad
E^\ddagger=u(x_s)=0.224613,
\qquad
\Lambda^2=-u''(x_s)=0.5918,
\label{eq:appendix-saddle-data}
\end{equation}
quoted to the digits used in the figures. For the outside-cusp comparison in Figure~\ref{fig:inside-outside}(b), we simply shift the asymmetry to $\beta=0.45$, for which only the single minimum at $x=-0.970980$ remains.

\textbf{Generating the cusp and lower-sheet contours (Figures~\ref{fig:cusp}--\ref{fig:cuts}).} Figure~\ref{fig:cusp} is entirely analytic. The upper and lower cusp boundaries, $\beta_\pm(\delta)=\pm(1-\delta^{2/3})^{3/2}$ for $0<\delta<1$, are plotted on a uniform mesh and shaded. Figures~\ref{fig:lower-sheet} and \ref{fig:cuts} evaluate the dimensionless lower sheet $v_-(x,\eta)=u(x)+\frac{\nu^2}{4}\eta^2$ and its 1D cut $v_-(x,0)$ on a dense rectangular grid, overlaying the critical points identified via Eq.~\eqref{eq:appendix-quartic}.

\textbf{Analytical expressions for the NHIM and invariant manifolds (Figures~\ref{fig:nhim-upo}--\ref{fig:manifolds}).} Because the chosen Hamiltonian \eqref{eq:appendix-H} is separable, the NHIM/UPO at energy $E=E^\ddagger+\Delta E$ can be written analytically without resorting to numerical shooting or differential correction. It is given exactly by:
\begin{equation}
x=x_s,
\qquad
p_x=0,
\qquad
\frac{1}{2} p_\eta^2+\frac{\nu^2}{4}\eta^2=\Delta E.
\label{eq:appendix-upo}
\end{equation}
In the center-mode plane, this forms the exact ellipse shown in Figure~\ref{fig:nhim-upo} for $\Delta E=0.01$, $0.03$, and $0.06$:
\begin{equation}
\eta(\theta)=\frac{2\sqrt{\Delta E}}{\nu}\cos\theta,
\qquad
p_\eta(\theta)=\sqrt{2\Delta E}\sin\theta.
\label{eq:appendix-upo-ellipse}
\end{equation}
At the saddle energy, the hyperbolic part of the Hamiltonian satisfies $\frac{1}{2} p_x^2+u(x)=E^\ddagger$. The zero-reactive-energy separatrix branches in the $(x,p_x)$ plane (Figure~\ref{fig:manifolds}) therefore satisfy:
\begin{equation}
p_x=\pm\sqrt{2\bigl(E^\ddagger-u(x)\bigr)}.
\label{eq:appendix-separatrix}
\end{equation}
Figure~\ref{fig:manifolds} plots Eq.~\eqref{eq:appendix-separatrix} directly over the full range $x_L\le x\le x_R$ between the turning points $x_L=-1.976$ and $x_R=1.166$ at which $u(x)=E^\ddagger$. No Poincar\'e map is needed for this separable benchmark: in the $(x,p_x)$ projection each complete homoclinic loop belongs to both the global stable and unstable manifolds. Colors mark approaching and departing portions, which coincide with the local stable and unstable branches near the saddle. 

\textbf{Constructing the dividing surface and trajectory sampling (Figures~\ref{fig:ds}--\ref{fig:inside-outside}).} For the representative inside-cusp parameter set with $\Delta E=0.03$, the closed forward and backward dividing-surface hemispheres on the energy shell $\mathcal{H}=E^\ddagger+\Delta E$ are given by:
\begin{equation}
\begin{gathered}
x=x_s,\qquad
\frac{1}{2}p_\eta^2+\frac{\nu^2}{4}\eta^2\leq\Delta E,\\
p_x=\pm\sqrt{2\left(\Delta E-\frac{1}{2}p_\eta^2-\frac{\nu^2}{4}\eta^2\right)}.
\end{gathered}
\label{eq:appendix-ds}
\end{equation}
Their common boundary at $p_x=0$ is exactly the NHIM/UPO; their interiors have $p_x>0$ and $p_x<0$, respectively. To illustrate the local sign of the crossing velocity (Figure~\ref{fig:ds}b), five equally spaced sample points $\eta_0\in0.75\,\eta_{\max}\{-1,-\tfrac12,0,\tfrac12,1\}$, with $\eta_{\max}=2\sqrt{\Delta E}/\nu$, are chosen on the line $p_\eta=0$ inside the forward dividing-surface ellipse. The conjugate momentum $p_x=\pm\bigl[2(\Delta E-\nu^2\eta_0^2/4)\bigr]^{1/2}$ is determined from the fixed-energy condition ($p_x>0$ for forward trajectories, $p_x<0$ for backward trajectories). Here forward and backward refer to crossing orientation; all ten trajectories are propagated forward in dimensionless time. Figure~\ref{fig:inside-outside}(a) shows the same ten trajectories. In Figure~\ref{fig:inside-outside}(b) no trajectories are launched: for the outside-cusp parameter set there is no lower-sheet saddle and hence no saddle-attached NHIM or associated local dividing surface, and the vertical line at $x=x_s$ is drawn only as a reference.

\textbf{Software, integration, and energy check.} All figures were generated with Python scripts, using \texttt{NumPy} for array operations and root finding, \texttt{SciPy} for the reference integration described below, and \texttt{Matplotlib} for visualization. The ten short trajectories shown in Figures~\ref{fig:ds}(b) and \ref{fig:inside-outside}(a) were integrated with a fixed-step classical fourth-order Runge--Kutta method with $\Delta t=0.01$ over the window $0\le t\le T=2.2$ (220 steps), which is approximately one quarter of the transverse period $2\pi\sqrt{2}/\nu$ and is the interval displayed in the figures. As a simple consistency check on this non-symplectic fixed-step integration, the Hamiltonian~\eqref{eq:appendix-H} was evaluated at every stored integration step: the maximum of $|\mathcal{H}(t)-\mathcal{H}(0)|$ over the 221 stored states of each of the ten trajectories is $8.7\times10^{-12}$, i.e.\ $3\times10^{-10}$ relative to $\Delta E=0.03$ and $4\times10^{-11}$ relative to $E^\ddagger$. Halving the step to $\Delta t=0.005$ changes the trajectory endpoints by less than $5\times10^{-11}$, and an adaptive eighth-order Dormand--Prince integration (\texttt{scipy.integrate.solve\_ivp}, method \texttt{DOP853}, relative tolerance $10^{-13}$ and absolute tolerance $10^{-15}$) agrees with the fixed-step endpoints to within $5\times10^{-11}$. Both endpoint comparisons use the maximum absolute component difference in $(x,\eta,p_x,p_\eta)$ at $t=T$, maximized over all ten trajectories. These trajectories serve only as local illustrations of the sign of the crossing velocity near the dividing surface. They are not used to estimate reactive flux, recrossing probabilities, gap times, or long-time transport, and the invariant objects themselves are obtained from the exact separable formulas above rather than validated by these trajectories.

\section*{Data and code availability}

The Python scripts that generate all figures in this paper, together with a diagnostics script that prints the quantities quoted in Appendix~\ref{app:numerics} (root-finding tolerances and residuals, saddle data, and the energy-conservation and step-halving checks for the short trajectories), are archived on Zenodo under the MIT License~\cite{Wiggins2026Code}. The package regenerates every figure from the model equations with a single command; it contains no random numbers or fitted data.

\section*{Declaration of generative AI and AI-assisted technologies in the manuscript preparation process}

During the preparation of this work, ChatGPT was used to assist with language editing, organization, checking of exposition, and the development and checking of numerical scripts and figures. The author takes full responsibility for the content of the published article.

\end{document}